\documentclass[%
reprint,
 amsmath,amssymb,
 aps,
]{revtex4-1}

\usepackage{graphicx}
\usepackage{dcolumn}
\usepackage{bm}
\usepackage[justification=raggedright]{caption}
\usepackage{subcaption}
\usepackage{physics}
\usepackage{algorithm}
\usepackage{algorithmic}
\usepackage{comment}
\usepackage{xcolor} 
\usepackage{hyperref}

\usepackage{scalerel,stackengine}
\stackMath
\newcommand\reallywidehat[1]{%
\savestack{\tmpbox}{\stretchto{%
  \scaleto{%
    \scalerel*[\widthof{\ensuremath{#1}}]{\kern.1pt\mathchar"0362\kern.1pt}%
    {\rule{0ex}{1\textheight}}
  }{0.7\textheight}%
}{2.4ex}}%
\stackon[-6.9pt]{#1}{\tmpbox}%
}
\newcommand{\W}{\bm{W}}
\newcommand{\dW}{d\W_t}
\newcommand{\delW}{\Delta\W}
\newcommand{\matI}{\bm{I}_3}

\newcommand{\vv}{\vb{v}}
\newcommand{\vnow}{\vb{v}_k}
\newcommand{\vnext}{\vb{v}_{k+1}}

\newcommand{\vmu}[1]{\bm{\mu}(#1)}

\newcommand{\vhat}{\check{\vv}}
\newcommand{\lm}[1]{#1_l^m}
\DeclareMathOperator{\Sgn}{Sgn}

\begin{document}

\preprint{APS/123-QED}

\title{Simulating Pitch Angle Scattering Using An Explicitly Solvable Energy Conserving Algorithm}

\author{Xin Zhang}
\author{Yichen Fu}
\author{Hong Qin}
\affiliation{Princeton Plasma Physics Laboratory}
\affiliation{Department of Astrophysical Sciences, Princeton University}

\date{\today}

\begin{abstract}


Particle distribution functions evolving under the Lorentz operator can be simulated with the Langevin equation for pitch angle scattering. This approach is frequently used in particle based Monte-Carlo simulations of plasma collisions, among others.
However, most numerical treatments do not guarantee energy conservation, which may lead to unphysical artifacts such as numerical heating and spectra distortions. We present a novel structure-preserving numerical algorithm for the Langevin equation for pitch angle scattering. Similar to the well-known Boris algorithm, the proposed numerical scheme takes advantage of the structure-preserving properties of the Cayley transform when calculating the velocity-space rotations. The resulting algorithm is explicitly solvable, while preserving the norm of velocities down to machine precision. We demonstrate that the method has the same order of numerical convergence as the traditional stochastic Euler-Maruyama method.
The numerical scheme is benchmarked by simulating the pitch angle scattering of a particle beam, and comparing with the analytical solution.
Benchmark results show excellent agreement with theoretical predictions, showcasing the remarkable long time accuracy of the proposed algorithm.
\end{abstract}

\maketitle

\section{\label{sec:intro}Introduction}




Coulomb collisions constitute one of the most basic forms of interaction among particles in a plasma. 
For each plasma scenario, simplifications can be made to the description of Coulomb collisions based on physical limits such as mass ratio and time scale ordering. 
One notable example is the Lorentz operator for pitch angle scattering. 
By itself, the Lorentz operator is frequently used to model electrons colliding with a cold stationary ion background; more generally, it appears as a term in the test particle collision operator for collisions with a stationary Maxwellian background \cite{liu2016collisionless, heikkinen1993monte, eriksson2003simulation, ichimaru2018statistical}. 
Since it is the limit of Coulomb collisions when the heavier particles are infinitely massive compared to the lighter particles, the Lorentz operator conserves particle energy.In a realistic plasma, this mass ratio is small but finite, and pitch angle scattering will result in an energy transfer of order $O(m/M)$, where $m$ and $M$ represent the mass of the lighter and heavier species respectively. Therefore, a numerical simulation in the zero mass ratio limit should not be allowed to introduce an energy error that is higher than $O(m/M)$. However, this energy conservation is not necessarily satisfied in numerical simulations.

A common technique to ``fix" the energy conservation is by recording the particle energy before the collisions, and then re-scaling the velocity vector after \cite{cadjan1999langevin, manheimer1997langevin}. Such techniques, while sufficient to some extent, are inherently \emph{ad hoc} and not ideal, because the choice of where to allocate the energy compensation is unspecified. For example, one can re-scale the magnitude of the vector while keeping its direction fixed, or choose to adjust an arbitrary component until energy is conserved. 
%
Alternatively, higher order stochastic integrators can be used \cite{rosin2014multilevel, dimits2013higher, cohen2010time}.
However, these methods only decrease the error in energy at each time step, instead of removing the error completely. 
They can also be computationally expensive, especially in multiple dimensions, if iterative root finding or sampling of correlated random processes is required \cite{dimits2013higher, d2006midpoint, rosin2014multilevel}.

One solution to overcome this difficulty is to adopt structure-preserving geometric algorithms analogous to those that have been successfully developed and applied for deterministic dynamical systems \cite{ruth83,feng1986difference,feng10,SanzSerna1988,sanz-serna94,hairer02,qin2008variational,he2017explicit,xiao2019a}. Existing work on structure-preserving stochastic algorithms mainly focus on symplectic stochastic systems \cite{milstein2002numerical, milstein2002symplectic, wang2007variational,wang2014generating,hong2017high,wang2017construction,zhou2017stochastic,holm2018stochastic}. However, these algorithms are not suitable when the system under study is not Hamiltonian, as is the case for the Lorentz operator. 


In this work, we present an energy conserving numerical scheme that explicitly advances the Langevin equation for pitch angle scatting. The energy conserving property is realized using the Cayley transform, which has been long adopted in deterministic differential equations to represent rotations, such as in the well-known Boris algorithm \cite{boris1970relativistic,qin2013boris} and other high-order volume-preserving algorithms \cite{he2015volume,he2016high,he2016higher}.
We show that the new algorithm is of global strong order $1/2$, similar to the classic Euler-Maruyama scheme, while particle energies are exactly conserved independent of time steps. We then benchmark the performance of the algorithm against an analytically solvable Fokker-Planck equation for the Lorentzian plasma, and demonstrate its excellent long time accuracy in both the calculated transport coefficients and the particle distribution functions. This is particularly important when the total time of the simulation exceeds a few collision times, as is the case with time-dependent simulations of a fusion plasma discharge. 

Although we focus on the Lorentz operator to highlight the structure-preserving properties, the new algorithm is in principle generalizable to non-energy conserving collisions. 
It is specifically applicable to particle-based simulations, and can be considered a complementary approach to directly solving the Fokker-Planck equations when a lighter weight calculation is desired. We note in passing that structure-preserving algorithms can also be applied to solve the Fokker-Planck equation, such as the recently proposed metriplectic methods for the Landau collision operators \cite{Kraus2017,hirvijoki2018}. These methods are excellent candidates for solving Fokker-Planck equations directly, and have distinct advantages when dealing with nonlinear systems.

The paper is organized as follows. Section~\ref{sec:Langevin} provides a brief review on the connections between the Fokker-Planck and the Langevin equations, and then derives the Langevin equation for pitch angle scattering. Section~\ref{sec:numerics} presents the explicitly solvable algorithm and studies its convergence behaviors. Section~\ref{sec:beam} introduces the benchmark problem on beam diffusion and shows the corresponding numerical results.

\section{\label{sec:Langevin} Background}
    


\subsection{Stochastic Differential Equations and the Fokker Planck Equation}
A simple example of a Langevin equation is the stochastic drag-diffusion equation known as the \emph{Ornstein-Uhlenbeck} process:
\begin{equation}\label{eq:lg_brown}
    \dv{x}{t} = -\nu x(t) + \tilde{a}(t),
\end{equation}
where $\nu>0$ is the constant that represents deterministic drag, and $\tilde{a}$ is the random variable describing the stochastic forcing. When representing Brownian motion, this stochastic forcing is interpreted as ``kicks" to the macroscopic particle by the thermal motion of the background particles \cite{uhlenbeck1930theory}. Assuming that:
\begin{align}
    \expval{\tilde{a}(t)} &= 0, \label{eq:a1}\\
    \expval{\tilde{a}(t)\tilde{a}(t')} &= 2D\delta(t - t'),\label{eq:a2}
\end{align}
where $\expval{\cdot}$ denotes expectation values,
the Langevin equation~(\ref{eq:lg_brown}) is equivalent to the Fokker Planck (FP) equation,
\begin{equation}\label{eq:FP_brown}
    \pdv{P(x, t)}{t} = \pdv{x}(\nu xP) + \pdv[2]{x}(DP),
\end{equation}
in the sense that the FP equation governs the transition probabilities $P(x, t)$ that the solutions $x(t)$ to equation (\ref{eq:lg_brown}) must satisfy  \cite{uhlenbeck1930theory, einstein1905motion, kloeden2013numerical}. 


The Langevin approach has earned great popularity in treating similar systems within the physical sciences, since it offers a more concrete picture with relatively small computational effort comparing to binary collision models \cite{takizuka1977binary}.
However, difficulties arise when the system under study responds nonlinearly to fluctuations, or, in other words, when the magnitude of the fluctuation depends on the state of the system itself:
\begin{equation}\label{eq:lg_nonlinear}
    \dv{x}{t} = -\nu x(t) + \tilde{a}(x, t).
\end{equation}
Since $\tilde{a}$ is only defined up to the second moment by Eq.~(\ref{eq:a1})-(\ref{eq:a2}), we are free to choose all higher moments. As is common in physics, we can choose the ``kicks" to be Gaussian distributed, both for mathematical convenience, and due to the fact that the cumulative statistics of many small random kicks is likely to be Gaussian based on the central limit theorem.

Before proceeding to discuss the nonlinear Langevin equation, we first transform Eq.~(\ref{eq:lg_nonlinear}) into the standard form of a stochastic differential equation (SDE):
\begin{equation}\label{eq:lg_wiener}
    dx = -\nu x(t)dt + \sigma(x, t)dW_t,
\end{equation}
where $\sigma(x, t)$ is now a deterministic function satisfying $\sigma^2(x, t) = 2D$, and $W_t$ denotes the standard Wiener process. Importantly, the increments of $W_t$ in time are independent of each other, and follow a Gaussian distribution with zero mean:
\begin{equation}\label{eq:wiener}
    \expval{W(t+\Delta t) - W(t)} \sim \mathcal{N}(0, \Delta t).
\end{equation}
The stochastic differential $dW_t$ in (\ref{eq:lg_nonlinear}) is then naturally defined as the $\Delta t \to 0$ limit of (\ref{eq:wiener}), also known as the Gaussian white noise. 

The Wiener process is a peculiar function that is continuous but nowhere differentiable. It can be understood as a Fourier series that includes all the frequencies:
\begin{equation}\label{eq:KL}
    W_t = \sum_{n=0}^{\infty} Z_n \phi_n(t) \text{   for } 0\leq t \leq T,
\end{equation}
where $Z_n$ are independent standard Gaussian random variables, and $\phi_n(t)$ are the usual Fourier basis functions normalized for the time interval $(0, T)$:
\begin{equation}
    \phi_n(t) = \frac{2\sqrt{2T}}{(2n+1)\pi}\sin\qty[\frac{(2n+1)\pi t}{2T}].
\end{equation}
This form is known as the \emph{Karhunen-Loeve expansion}, a truncated version of which can be a convenient method for numerical implementation \cite{kloeden2013numerical}.

Returning to the SDE (\ref{eq:lg_wiener}), we can now intuitively interpret the fluctuation term as a series of kicks whose magnitude is a Gaussian random number scaled by the factor $\sigma(x, t)$. However, a question remains: when during the time interval $(t, t+dt)$ does the kick arrive? Since $W_t$ is nowhere differentiable, this choice in interpretation leads to distinct solutions. 

This problem has now been coined as the \emph{Ito-Stratonovich dilemma} \cite{van2007stochastic}, named after the two popular interpretations of stochastic calculus. In the \emph{Ito} interpretation, all kicks arrive at the beginning of the time interval, whereas in the Stratonovich interpretation, each discrete kick is understood as the \emph{average} forcing within the (infinitely narrow) time interval. This dilemma arises whenever the stochastic differential $dW$ is multiplied by a non-constant function, a situation frequently termed as \emph{multiplicative noise}. The converse is termed \emph{additive noise}. SDE systems with multiplicative noise is frequently seen in plasma physics and beyond. When treating these systems, one must take care in choosing the proper interpretation. Although both are mathematically valid, the wrong choice could lead to invalid physical results \cite{van2007stochastic}.

Similar to the equivalency between equations (\ref{eq:lg_brown}) and (\ref{eq:FP_brown}), it has been shown through stochastic calculus that, in general, the vector SDEs:
\begin{align}
    \text{Ito:  }&d\vv = \bm{\mu}(\vv)dt + \bm{\sigma}(\vv)\dW
    \label{eq:SDE_ito_0},\\
    \text{Stratonovich:  }& d\vv = \bm{\bar{\mu}}(\vv)dt + \bm{\sigma}(\vv)\circ\dW,
    \label{eq:SDE_strat_0}
\end{align}
(where $\circ$ denotes Stratonovich calculus) are equivalent to the FP equations in the Ito form, \begin{equation}\label{eq:FP_Ito}
    \pdv{f(\vv, t)}{t} = -\pdv{\vv}\cdot\bm{\mu}(\vv) f + \pdv{\vv}\pdv{\vv}:\bm{D}(\vv)f,
\end{equation}
and the Stratonovich form:
\begin{equation}\label{eq:FP_Strat}
    \pdv{f(\vv, t)}{t} = -\pdv{\vv}\cdot\bm{\bar{\mu}}(\vv) f + \frac{1}{2}\pdv{\vv} \cdot\bm{\sigma}(\vv)\cdot \pdv{\vv}\cdot(\bm{\sigma}(\vv)f),
\end{equation}
respectively, where we have defined $\bm{D} = \frac{1}{2}\bm{\sigma}\bm{\sigma}^T$ \cite{ottinger2012stochastic}. These two forms of the FP equations will in general have different drag coefficients when the diffusion tensor $\bm{D}$ is a function of $\vv$. Note that because the diffusion tensor $\bm{D}$ is required to be positive semi-definite \cite{van2007stochastic}, the decomposition of $\bm{D}$ into $\frac{1}{2}\bm{\sigma}\bm{\sigma}^T$ is in general possible, for example, via Cholesky decomposition \cite{press1988numerical}. 


\subsection{The Langevin Equations for Pitch Angle Scattering}
The FP equation for pitch angle scattering is:
\begin{equation}\label{eq:Cei}
\pdv{f_e(\vv, t)}{t} = \frac{1}{2}\hat{\mu}_{ei}\pdv{\vv}\cdot \bm{U}(\vv)\cdot \pdv{f_e}{\vv},
\end{equation}
where $\bm{U}(\vv) \equiv \frac{1}{v}(\matI - \vhat\vhat)$, $\vhat$ is the unit vector in $\vv$ direction \cite{ichimaru2018statistical}. The right hand side of Eq.~\ref{eq:Cei} is the Lorentz operator. The constant $\hat{\mu}_{ei}$ is given by:
\begin{align}
\hat{\mu}_{ab} = \frac{n_b q_a^2 q_b^2 \ln\Lambda }{4\pi\epsilon_0^2 m_a^2}
	= \nu_{ab}v_a^3,
\end{align}
where $\nu_{ei}$ is the standard thermal collision frequency \cite{huba2007nrl}. The energy conservation of the system can be easily demonstrated by integrating against $v^2$, taking advantage of the fact that $\bm{U}$ projects onto the plane perpendicular to $\vv$.

In order to find a Langevin SDE whose statistical ensemble reproduces the behavior described by the FP equation~(\ref{eq:Cei}), we need to first transform the FP equation into the Ito form (\ref{eq:FP_Ito}) and the Stratonovich form (\ref{eq:FP_Strat}). Through straightforward algebraic manipulations, we find that for the given the FP equation~(\ref{eq:Cei}), the drag and diffusion coefficients are
\begin{align}
    \bm{\mu}(\vv) &= -\hat{\mu}_{ei}\vv/v^3,\\
    \bm{\bar{\mu}}(\vv) &= 0,\\
    \bm{\sigma}(\vv) &= \bm{\sigma}^T(\vv) = \sqrt{\frac{\hat{\mu}_{ei}}{v}}(\matI - \vhat\vhat),
\end{align}
in the notations consistent with equations (\ref{eq:FP_Ito}) and (\ref{eq:FP_Strat}). 
We then arrive at the final Langevin equations that will be solved numerically in the rest of the paper:
\begin{align}
    \text{Ito:  }&d\vv = -\hat{\mu}_{ei}\frac{\vv}{v^3}
    dt + \sqrt{\frac{\hat{\mu}_{ei}}{v}}(\matI - \vhat\vhat)\dW, \label{eq:SDE_Ito}\\
    \text{Stratonovich:  }& d\vv =  \sqrt{\frac{\hat{\mu}_{ei}}{v}}(\matI - \vhat\vhat)\circ\dW. \label{eq:SDE_Strat}
\end{align}
Despite their appearances, these two equations are mathematically equivalent, and lead to the same solution. They are both still energy conserving when integrated with the correct choice of stochastic calculus. 

Equations of a similar structure are also seen whenever an SDE is desired to simulate the effect of the Lorentz operator, for example, in the stochastic Landau-Lifshitz dynamics of magnetization \cite{d2006midpoint}. The algorithm that we proceed to derive in the next sections will also be suitable for such equations outside of plasma physics when norm-preservation is desired.

\subsection{Ito-Stratanovich Dilemma and Numerical Methods for SDEs}

    
    
The \emph{Ito-Stratonovich dilemma} in the interpretation of stochastic calculus has interesting consequences for numerical algorithms. Specifically, each choice of discretization may inherently correspond to one type of interpretation, while being completely incompatible with the other. The stochastic generalizations of the forward Euler method and the midpoint method, for example, respectively correspond to the Ito and Stratonovich interpretations. 
In this section we will briefly review both of these methods as they are closely related to the proposed new algorithm. For simplicity of notations, we set the constant $\hat{\mu}_{ei}\equiv1$ for the rest of this section. 

The popular Euler-Maruyama (EM) method (Alg.~\ref{alg:EM}), is the natural generalization of the deterministic forward-Euler method to stochastic calculus, where each increment in time is advanced with the \emph{current} derivative \cite{kloeden2013numerical}. This directly aligns with the Ito interpretation, where the stochastic kicks come in at the beginning of the time interval. 
Observing that the projection operator $(\matI - \check{\vv}\check{\vv})$ could be written as two cross products:
\begin{align}
(\matI - \check{\vv}\check{\vv})\dW=(\check{\vv}\times \dW)\times\check{\vv},
\label{eq:cross_product}
\end{align}
the Euler-Maruyama scheme for Eq.~(\ref{eq:SDE_Ito}) is given by \cite{kloeden2013numerical}:
\begin{align}
\bar{\vv}_{k+1}^{EM} - \vv_k = 
-\dfrac{\vv_k}{v_k^3} h + 
\dfrac{(\vv_{k}\times\Delta{\W})\times {\vv}_{k}}{v_{k}^{5/2}},
\label{eq:numerical_SDE_EM}
\end{align}
where $h$ is the step size in time, $v_k=\norm{\vv_k}$ is the norm of the velocity,  and $\Delta \W = \W(t+h) - \W(t)\sim \mathcal{N}(0,\matI h)$ is a vector Wiener process. The EM scheme is fully explicit, similar to their deterministic counterpart, and is therefore straightforward to implement. We stress that the EM method strictly correspond to the Ito interpretation, and at the continuous limit the energy is conserved. However, one can show that the norm of the velocities $v_k$ is not preserved with finite time-step $h$. Integrating Eq.~(\ref{eq:SDE_Strat}) with the stochastic EM method, for example, will lead to a catastrophic energy error  that is unbounded in time (see Appendix~\ref{sec:app}).

\begin{algorithm}[H]
    \caption{The Euler-Maruyama (EM) Method}
    \label{alg:EM}
    \algsetup{indent=2em}
    \begin{algorithmic}[1]
        \REQUIRE Initial velocity $\vv_0$, time interval $T$, step size $h$
        \REQUIRE A prescribed sample path $\W(t)$
        
        \FOR{$k=0$ \TO $T/h$}
            \STATE $t_k = t_0 + k h$
            \STATE $\delW = \W(t_k+h) - \W(t_k)$ 
            \STATE $\vnext = \vnow + \bm{\mu}(\vnow)h+ \bm{\sigma}(\vnow)\delW$
        \ENDFOR
    \end{algorithmic}
\end{algorithm}

Similarly, the Stratanovich interpretation naturally correspond to implicit methods of integration, where information at both the beginning and the end of the time interval is required. A classic example is the stochastic midpoint method, where the function is advanced with the \emph{average} of the derivatives at both end points \cite{milstein2002numerical, milstein2013stochastic}. The midpoint discretization for Eq.~(\ref{eq:SDE_Strat}) therefore reads:
\begin{align}
\bar{\vv}^{MP}_{k+1} - \vv_k = \frac{(\bar{\vv}_{k+1/2} \times \Delta \W)\times \bar{\vv}_{k+1/2}}{\bar{v}_{k+1/2}^{5/2}}, 
\label{eq:numerical_SDE_MD}
\end{align}
where $\bar{\vv}_{k+1/2}\equiv (\vv_k + \bar{\vv}^{MP}_{k+1}) / 2$ is the velocity at the midpoint, and  $\bar{v}_{k+1/2} = \norm{\bar{\vv}_{k+1/2}}$ is the norm of the velocity vectors. As pointed out in Ref.~\cite{d2006midpoint}, the midpoint scheme preserves the magnitude of velocity in principle. However, $\bar{\vv}_{k+1}$ can not be explicitly solved from Eq.~(\ref{eq:numerical_SDE_MD}). This means that a root finding routine such as the Newton iteration is required to solve for the midpoint $\bar{\vv}_{k+1}$ at each time step, and the resulting accuracy of the velocity magnitude $v_k$ depends on the convergence of the root finding \cite{d2006midpoint, d2005geometrical}. Moreover, like any implicit integrator, the necessity of root finding at each time step adds significantly to the total computational cost.


\section{\label{sec:numerics} The Explicitly Solvable Energy Conserving Algorithm}
We propose the following implicit discretization for the Stratonovich SDE (\ref{eq:SDE_Strat}):
\begin{align}
\bar{\vv}_{k+1}^{ES} - \vv_k = \frac{({\vv}_{k} \times \Delta \W)\times \bar{\vv}_{k+1/2}}{{v}_{k}^{5/2}}, 
\label{eq:numerical_SDE_ES}
\end{align}
where again $\bar{\vv}_{k+1/2}\equiv (\vv_k + \bar{\vv}_{k+1}^{ES}) / 2$. 
Before proceeding to demonstrate the numerical convergence of the proposed algorithm, we first solve for $\bar{\vv}_{k+1}$ explicitly as promised. This is possible because the dependency on the future state $\bar{\vv}_{k+1}$ is linear on both sides of the equation. Since the cross product between two vectors $\vb{X}, \vb{Y}$ could be written as the product of the skew-symmetric matrix $\hat{\vb{X}}$ and the vector $\vb{Y}$,
\begin{align}
\vb{X} \times \vb{Y} \equiv \hat{\vb{X}} \vb{Y} :=
\begin{pmatrix}
0 & -X_3 & X_2 \\
X_3 & 0 & -X_1 \\
-X_2 & X_1 & 0
\end{pmatrix}
\begin{pmatrix} Y_1 \\ Y_2 \\ Y_3 \end{pmatrix}.
\label{eq:skew}
\end{align}
We can define a skew-symmetric matrix $\hat{\vb{M}}_k$ from vector $\vb{M}_k$:
\begin{align}\label{eq:cross}
\vb{M}_k := 
\dfrac{\vv_k \times \Delta \W}{2 v_k^{5/2}},
\end{align}
which depends only on the current state $\vv_{k}$. Then $\bar{\vv}_{k+1}^{ES}$ is explicitly solved by:
\begin{align}
\bar{\vv}_{k+1}^{ES} = \mathcal{C}(\hat{\vb{M}}_k) \vv_k,
\end{align}
where 
\begin{equation}\label{eq:cayley}
    \mathcal{C}(\hat{\vb{M}}_k):=(1-\hat{\vb{M}}_k)^{-1}(1+\hat{\vb{M}}_k)
\end{equation} 
is the Cayley transform of matrix $\hat{\vb{a}}_k$ \cite{feng1986difference,qin2013boris}. The Cayley transform can be numerically computed either with direct matrix inversion, or with a Rodriguez-type formula \cite{piggott2016geometric}. Since the matrix $\hat{\vb{M}}_k$ is skew-symmetric, an explicit formula for the Cayley transform could be derived (see Appendix~\ref{sec:explicit_cayley}). The algorithm is summarized in Alg.~\ref{alg:ES}.

\begin{algorithm}[H]
    \caption{The Explicitly Solvable (ES) Energy Conserving Method}
    \label{alg:ES}
    \algsetup{indent=2em}
    \begin{algorithmic}[1]
        \REQUIRE Initial velocity $\vv_0$, time interval $T$, step size $h$
        \REQUIRE A prescribed sample path $\W(t)$ 
        \FOR{$k=0$ \TO $T/h$}
            \STATE $t_k = t_0 + kh$
            \STATE $\delW = \W(t_k+h) - \W(t_k)$ 
            \STATE $\vb{M}_k = \vv_k \times \Delta \W / 2 v_k^{5/2} $
            \STATE $\mathcal{C}(\hat{\vb{M}}_k)= (1-\hat{\vb{M}}_k)^{-1}(1+\hat{\vb{M}}_k)$
            \STATE $\vnext = \mathcal{C}(\hat{\vb{M}}_k)\vnow$
        \ENDFOR
    \end{algorithmic}
\end{algorithm}
The conservation of energy can be easily verified by dotting both side of Eq.~(\ref{eq:numerical_SDE_ES}) with $\bar{\vv}_{k+1/2}$, which gives $v_k^2-\bar{v}_{k+1}^2=0$. 

\subsection{Strong and Weak Convergence of Numerical Errors
\label{sec:Strong_and_weak}}

Similar to the truncation errors in deterministic numerical schemes, the strong and weak errors of stochastic numerical schemes are central to understanding its convergence properties \cite{kloeden2013numerical, dimits2013higher}. In this section we will first define strong and weak errors, and then argue that the proposed explicitly solvable (ES) algorithm has the same order of convergence as both the EM method and the midpoint method.

For SDEs (\ref{eq:SDE_ito_0}) and (\ref{eq:SDE_strat_0}) with initial condition $\vv=\vv_0$ at $t\in [t_0, T]$, the definition of global strong and weak error for the time interval is given by:
\begin{align}
\epsilon_\text{s} &:=  \langle\left|\vv(T;t_0,\vv_0)-\bar{\vv}(T;t_0,\vv_0, h)\right|\rangle, \label{eq:strong}\\
\epsilon_\text{w} &:= \left| \langle\vv(T;t_0,\vv_0) - \bar{\vv}(T;t_0,\vv_0, h)\rangle \right|,
\label{eq:weak}
\end{align}
where $\vv(t;t_0,\vv_0)$ is the exact solution, $\bar{\vv}(t_k;t_0,\vv_0, h)$ is the time discrete approximation, and $t_k = t_0 + kh$. As usual, $\langle\cdot\rangle$ denotes expectation values, and $|\cdot|$ denotes absolute values. Intuitively, the strong error measures the errors of individual sample paths, whereas the weak error measures the error of the statistics distribution.

A numerical scheme is said to converge strongly with order $\alpha$ and weakly with order ${\beta}$, if there exists finite and independent constants $C_1$ and $C_2$, and a positive constant $h_0$, such that
\begin{align}
    &\epsilon_s \leq C_1 h^{\alpha}\text{, and}\\
    &\epsilon_w \leq C_2 h^{\beta},
\end{align}
for any $h \in (0, h_0)$ \cite{kloeden2013numerical}. Both the EM scheme and the midpoint scheme are of strong order $1/2$ and weak order $1$ \cite{kloeden2013numerical, milstein2002numerical}. 

A closely related idea to the above stated global strong and weak error is the concept of one-step strong and weak errors:
\begin{align}
\epsilon_\text{s,o} &:= \langle\left|\vv(t_0+h; t_0,\vv_0)-\bar{\vv}(t_0+h; t_0,\vv_0,h)\right|   \rangle; \\
\epsilon_\text{w,o} &:= \left| \langle \vv(t_0+h; t_0,\vv_0) - \bar{\vv}(t_0+h; t_0,\vv_0,h) \rangle\right|.
\end{align}
For a given numerical algorithm with one-step errors of order $p_1$ in the weak sense and $p_2$ in the strong sense, the algorithm is known to convergence globally with strong order $p_2-1/2$ if and only if $p_2\leq 1/2, p_1\leq p_2+1/2$ \cite{milstein2013stochastic}. Taking advantage of this fact, we found that the proposed ES algorithm converges strongly with global error of order $1/2$, which also implies that the algorithm converges weakly globally as well. Details of this calculation are included in Appendix~\ref{sec:proof}. A more rigorous proof of convergence and detailed discussions on the numerical properties of the algorithm will be included in a separate article \cite{Fu2020}. 



\subsection{Numerical verification of convergence}


To examine the convergence of strong and weak errors numerically, the definitions (\ref{eq:strong}) and (\ref{eq:weak}) are not feasible since the analytical solutions of the SDEs are unknown. However, we could define the following relative errors for time step $h_l$:
\begin{align}
\bar{\epsilon}_{\text{s},l} &= \expval{\left| \vv_\omega(T; h_{l+1}) - \vv_\omega(T; h_{l}) \right|}, \\
\bar{\epsilon}_{\text{w},l} &=  \left| \expval{\vv_\omega(T; h_{l+1})}  - \expval{\vv_\omega(T; h_{l})} \right|, 
\end{align}
where $\expval{\cdot}$ denotes ensemble average. It is easy to see that for an algorithm with strong order $\alpha$ and weak order $\beta$, these definitions of strong and weak errors converge at the same rate: $\bar{\epsilon}_\text{s,l}\sim\mathcal{O}(h_l^\alpha)$, and $\bar{\epsilon}_\text{w,l}\sim\mathcal{O}(h_l^\beta)$. 

\begin{algorithm}[H]
    \caption{Strong Convergence Test}
    \label{alg:strConv}
    \algsetup{indent=2em}
    \begin{algorithmic}[1]
        \REQUIRE Time interval $T$, initial velocity $\vv_0$
        \REQUIRE $\Omega$ independent Wiener processes $\W_\omega(t)$
        \REQUIRE Number of discretization levels $L$
        \REQUIRE Test algorithm $\bm{Alg}$
        \FOR{$\omega = 0$ \TO $\Omega$}
            \FOR{$l = 0$ \TO $L$}
                \STATE Calculate step size $h_l = T / 2^{l}$
                \STATE Find $\vv_\omega(T; h_l)$ by $\bm{Alg}$ using Wiener process $\W_\omega$ and time step $h_l$
                \IF{$l > 0$}
                    \STATE $\delta\vv_{\omega}(h_{l-1}) = |\vv_\omega(T; h_{l}) - \vv_\omega(T; h_{l-1})|$
                \ENDIF
            \ENDFOR
        \ENDFOR
    \end{algorithmic}
\end{algorithm}

\begin{algorithm}[H]
    \caption{Weak Convergence Test}
    \label{alg:weakConv}
    \algsetup{indent=2em}
    \begin{algorithmic}[1]
        \REQUIRE Time interval $T$, initial velocity $\vv_0$
        \REQUIRE $\Omega$ independent Wiener processes $\W_\omega(t)$
        \REQUIRE Number of discretization levels $L$
        \REQUIRE Test algorithm $\bm{Alg}$
        \FOR{$l=0$ \TO $L$}
            \FOR{$\omega = 0$ \TO $\Omega$}
                \STATE Calculate step size $h_l = T / 2^{l}$
                \STATE Find $\vv_\omega(T; h_l)$ by $\bm{Alg}$ using Wiener process $\W_\omega$ and time step $h_l$
            \ENDFOR
            
            \STATE Calculate $\vmu{h_l} = \expval{\vv_{\omega}(T; h_l)}$ 
            \IF{$l > 0$}
                \STATE $\delta \vmu{h_{l-1}} = \vmu{h_{l}} - \vmu{h_{l-1}}$
            \ENDIF
        \ENDFOR
    \end{algorithmic}
\end{algorithm}

\begin{figure}[ht]
    \centering
    \includegraphics[width=0.8\linewidth]{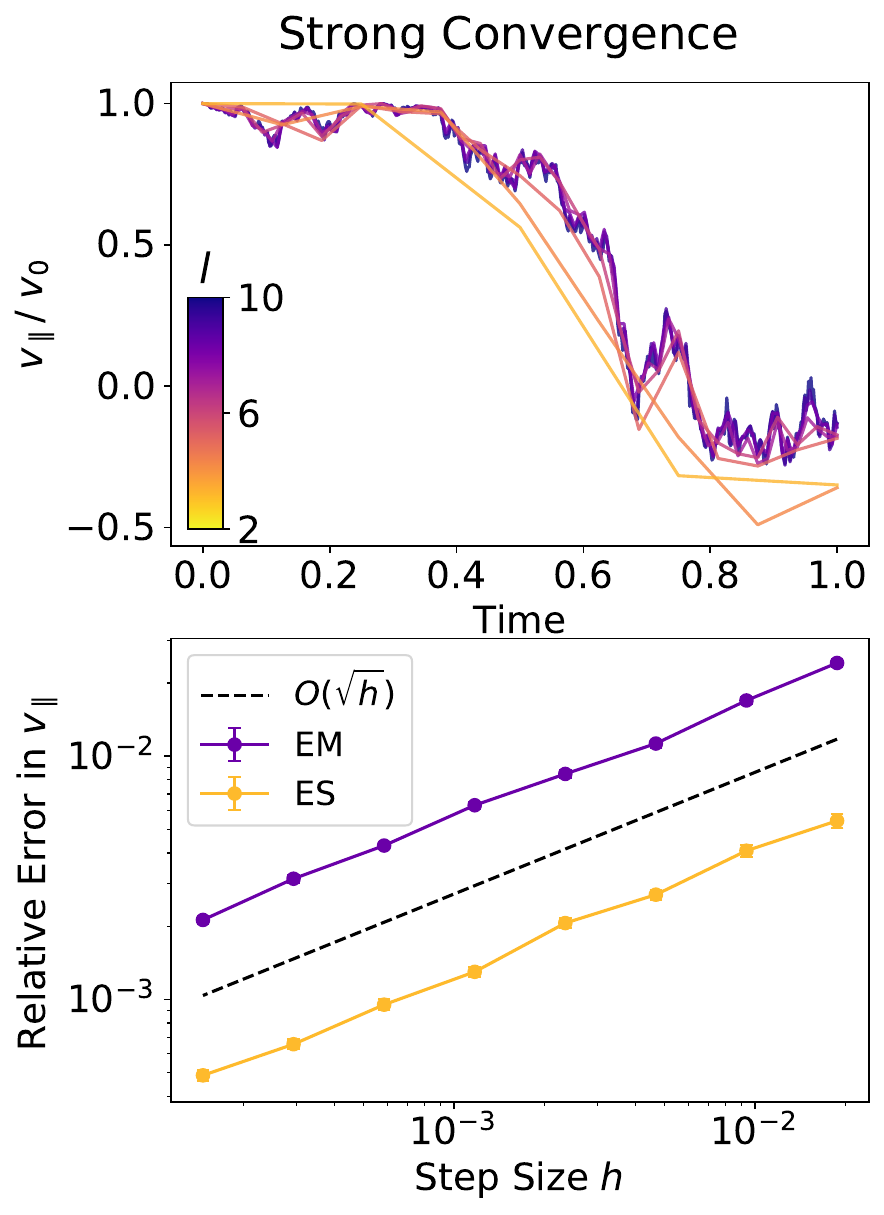}
    \caption{Global strong convergence. (top) Example solutions for one underlying Wiener process $\W(t)$ with increasing discretization levels $l$, calculated with the ES method. The solutions converge as time step size $h = 2^{-l}$ approaches zero. Color of lines corresponds to the values of $l$. (bottom) Scaling of global strong error at $t=T$ with step size $h$ shows a clear convergence rate of order $1/2$ (error bars too small to be visible), same for both the EM and ES method. A reference line for the expected convergence rate is shown as black dashed line. Sample size $N = 10^3$.}
    \label{fig:strongConv}
\end{figure}

For numerical tests of global strong convergence, the Wiener processes are prepared with the Karhunen-Loeve expansion given in Eq.~(\ref{eq:KL}). The discrete approximations are then found with different time step sizes $h_l=T / 2^{l}$, where $l$ denotes discretization level, and the numerical errors are computed at the end of the time interval $t=T$. The detailed procedures for the strong convergence test are given in Alg.~\ref{alg:strConv}. Figure~\ref{fig:strongConv} shows the numerical results of the strong convergence test, with total computation time $T$ normalized to $1$. The top panel shows one set of approximate solutions for a single underlying Wiener process, with different discretization levels. 
We specifically show the convergence of $v_\parallel$ as an example, defined as the component of $\vv$ parallel to the initial condition $\vv_0$. The sample paths of $v_{\parallel}(t;h_l)$ clearly converge as $h_l$ approaches zero. The bottom panel shows the strong convergence of global error for both the EM method and the ES method. Comparing with the reference line for $\mathcal{O}(\sqrt{h})$, both the EM and ES methods show a clear global strong convergence of order $1/2$, consistent with expectations. 


\begin{figure}[ht]
    \centering
    \includegraphics[width = 0.8\linewidth]{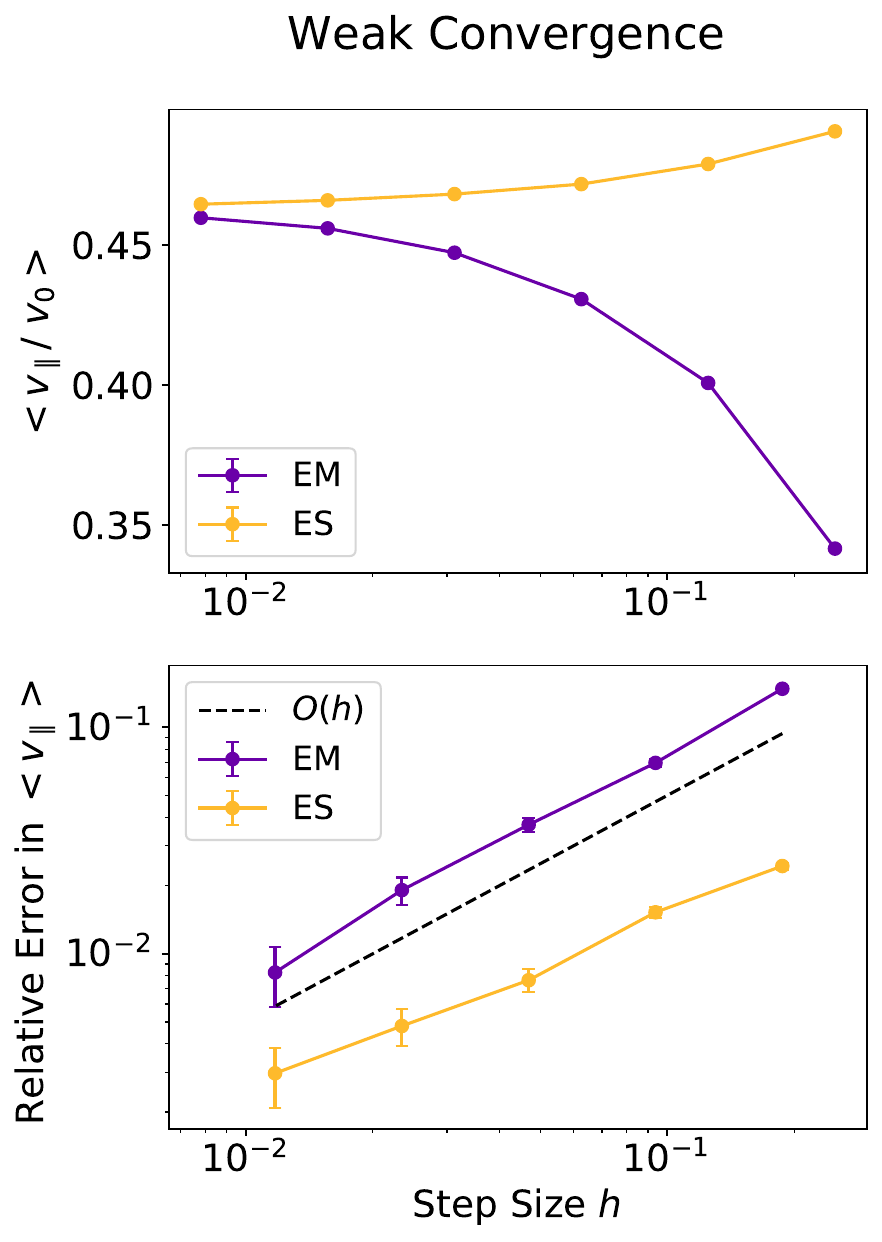}
    \caption{Global weak convergence. (top) Convergence of calculated $\expval{v_\parallel}$ as step size $h$ approaches zero. (bottom) Scaling of relative error with step size $h$ shows an approximate order 1 convergence. Dashed lines connect points for each method to guide the eye. A reference line for the expected order 1 is shown in black dashed line. Sample size $N = 10^9$.}
    \label{fig:weakConv}
\end{figure}

For the weak convergence tests, the underlying Wiener processes are regenerated for each individual sample path, and the ensemble averages are calculated at the end of the computational interval. The detailed procedures for the weak convergence tests are given in Alg.~\ref{alg:weakConv} and the numerical results are shown in Fig.~\ref{fig:weakConv}. The ensemble average of $v_\parallel$ shows clear signs of global convergence as $h_l$ approaches zero with convergence rate similar to that of the EM method.

\begin{figure}[ht]
    \centering
    \includegraphics[width = 0.8\linewidth]{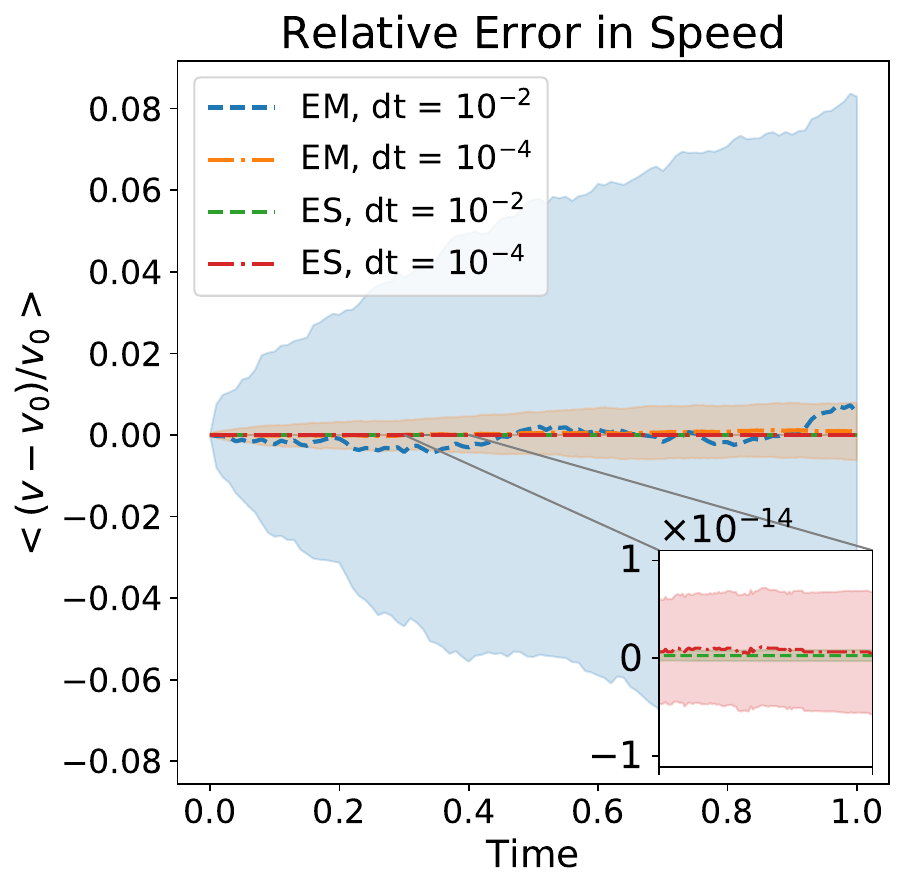}
    \caption{Comparison of the energy conserving properties of the proposed explicitly solvable algorithm and the Euler-Maruyama method. The lines shows the average speed of the ensemble, and the shaded area shows the spread within the ensemble. Sample size $N = 10^3$.}
    \label{fig:speed_rel}
\end{figure}

Figure~\ref{fig:speed_rel} shows the promised energy conservation properties of the ES method, compared with the EM method. The average particle speeds are shown as lines with shades showing the standard deviation within the ensemble. Particle speeds calculated from the EM method shows significant spread even with a very small time stepsize $10^{-4}$. In contrast, the error in particle speeds from the ES method remains close to the machine precision even for larger stepsizes.

\section{Beam Diffusion in Velocity Space}\label{sec:beam}
The setup of the benchmarking problem is as follows. A collection of $N$ charged particles is injected into a background Maxwellian plasma at uniform initial velocity $\vv_0$. According to Eq.~(\ref{eq:Cei}), the evolution of the test particle distribution function can be written as:
\begin{equation}\label{eq:beam}
    \pdv{f(\vv, t)}{t} = C_{ei}[f] + S,
\end{equation}
where $C_{ei}$ is the pitch angle scattering operator defined in (\ref{eq:Cei}) and $S$ is the source function given by:
\begin{equation}
    S = N\delta(\vv - \vv_0)\delta(t).
\end{equation}
This test problem is analytically solvable by noting that, in spherical coordinates,
\begin{equation}
    C_{ei}[f] = -\nu_{ei}\mathcal{L}[f],
\end{equation}
where $\mathcal{L}$ is the Lorentz operator
\begin{equation}
    \mathcal{L} := - \qty(\frac1{\sin\theta}\pdv{\theta} \sin\theta \pdv \theta
    + \frac1{\sin^2 \theta}\pdv[2]{\phi}).
\end{equation}
We can then find the series solution to (\ref{eq:beam}) as
\begin{equation}\label{eq:f_series}
   f(v, \theta, \phi) = \sum_{l=0}^{\infty}\sum_{m=-l}^{l}\lm{f}(v)\lm{Y}(\theta, \phi) ,
\end{equation}
where $\lm{Y}$ are the eigen functions of the Lorentz operator, known as the spherical harmonics:
\begin{align}\label{eq:eigenvalues}
&\mathcal{L}[\lm{Y}(\theta, \phi)] = l(l+1)\lm{Y}(\theta, \phi),\\
&\lm{Y}(\theta, \phi) \equiv \lm{P}(\cos \theta)\exp{im\phi}.
\end{align}
and $P_l^m$ is the associated Legendre functions of the first kind. 

Taking advantage of the ortho-normality of the Legendre series, the differential equation can be solved term by term in $l$. The exact form of the series coefficients $\lm{f}(v)$ can be found through fairly straightforward calculations, giving the final solution for the dynamics of the beam distribution:
\begin{align}\label{eq:f_solved}
f(\vv, t) = 
    &\frac{N\Theta(t)}{2\pi v_0^2}\delta(v - v_0)\times \nonumber\\
	&\sum_{l=0}^{\infty}\qty(\frac{2l+1}{2})e^{- \nu_{ei} l(l+1)t/2}P_l\qty(\frac{v_\parallel}{v_0}),
\end{align}
where $\Theta(t)$ is the Heaviside step function, and $v_\parallel$ is the velocity parallel to the beam initial velocity 
\begin{equation}
    v_\parallel \equiv (\vv\cdot\vv_0)/v_0.
\end{equation}
Without loss of generality, the initial beam axis can be aligned with the $x$ axis: \begin{align}
    \vv_0 &= v_{x0},\\
    v_\parallel &= v\sin\theta,\\
    |v_\perp| &= \sqrt{v_y^2 + v_z^2} = v\cos\theta \text{, and}\\
    v_y/v_z &= \tan{\phi}.
\end{align} 
This definition of the coordinate system is adopted in all figures in the current section.
\begin{figure}
    \centering
    \includegraphics[width = 0.8\linewidth]{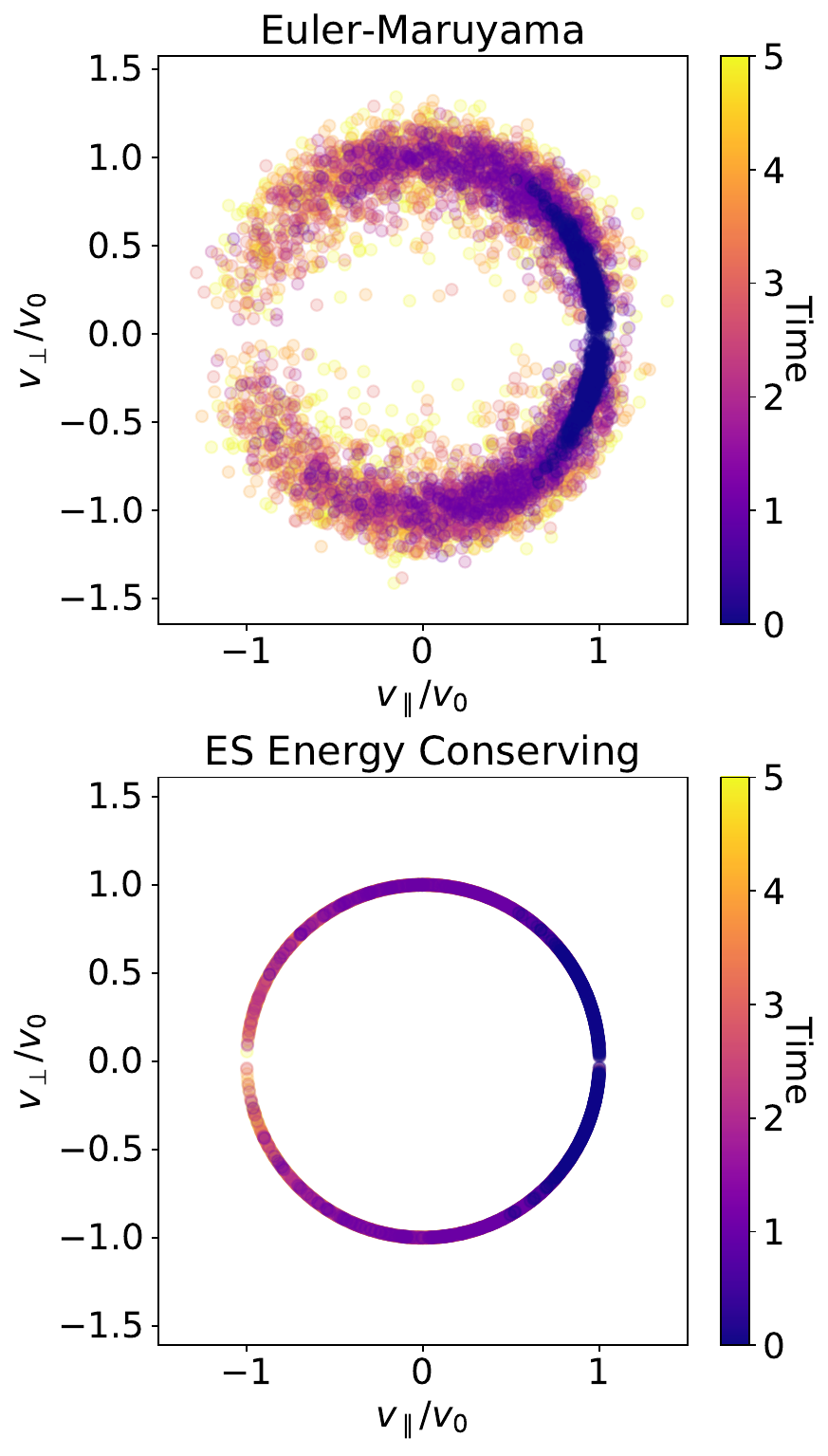}
    \caption{Comparison of velocity space rotation calculated with the EM method (top) and the ES energy conserving algorithm (bottom). Velocity space location is shown as $v_\perp = \Sgn(v_y)\sqrt{v_y^2 + v_z^2}$ v.s. $v_\parallel = v_x$. While EM method produces considerable spread in paricle energy, the ES method conserves energy exactly. Sample size $N = 10^3$.}
    \label{fig:circle}
\end{figure}

A few physical insights can be gained from the analytical solution. First of all, the particle speed (energy) is indeed conserved, since the dependency of $f(\vv, t)$ on the magnitude of velocity is a delta function $\delta(v - v_0)$ at the initial speed. Second, the distribution only depends on the pitch $v_\parallel / v_0 = \sin\theta$ of the particles, and not on the azimuthal phase $\phi$. Since the initial condition is azimuthally symmetric, this symmetry will also be preserved when evolving in time. In other words, our physical system is confined to evolve along ``rings" on a unit sphere, with two ignorable coordinates velocity magnitude $v$ and aximuthal phase $\phi$, and only one degree of freedom $v_\parallel/v = \sin\theta$.

\begin{figure}[h]
    \centering
    \begin{subfigure}[t]{0.8\linewidth}
        \centering
        \includegraphics[width=\linewidth]{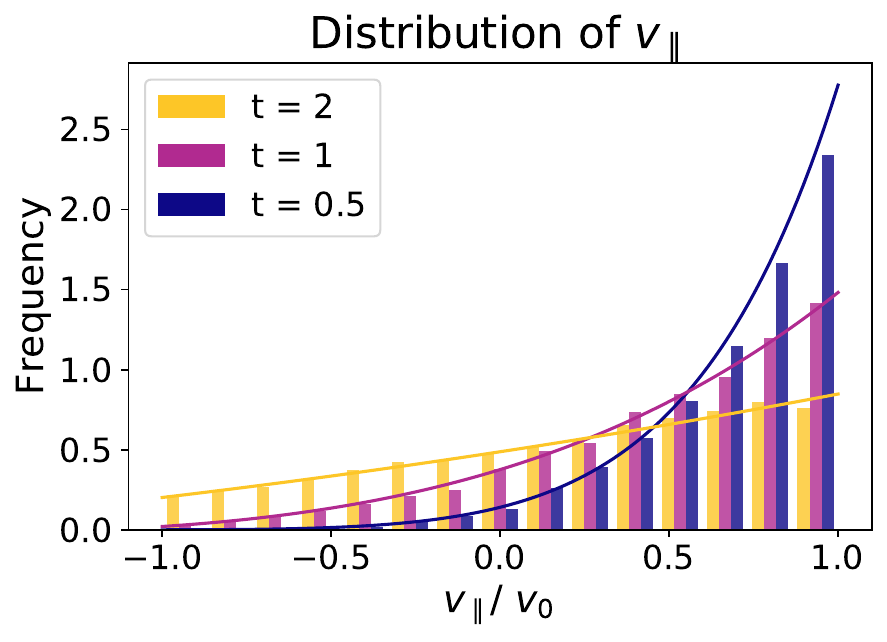}
    \end{subfigure}%
    
    \begin{subfigure}[t]{0.8\linewidth}
        \centering
        \includegraphics[width=\linewidth]{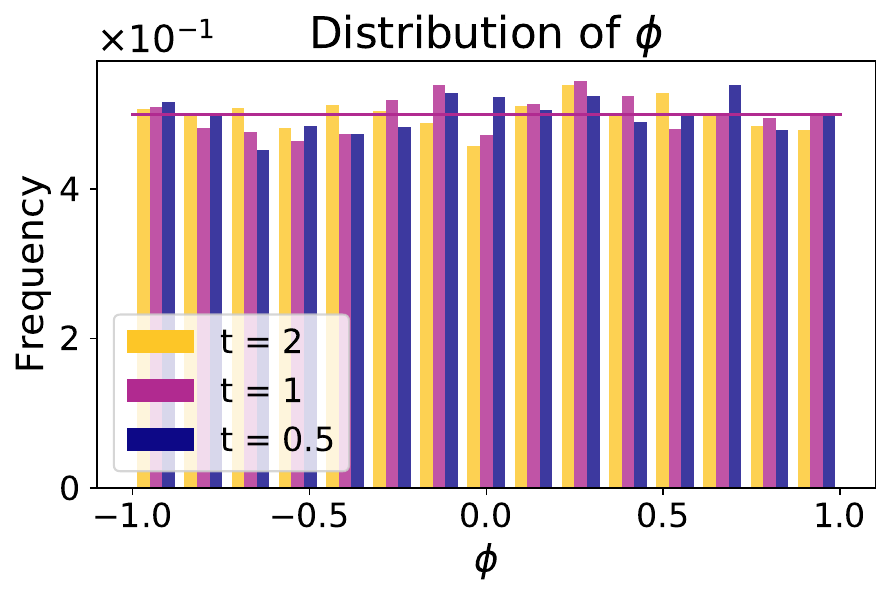}
    \end{subfigure}
    \caption{Simulated distributions of $v_{\parallel}/v_0$ and azimuthal angle $\phi = \frac2{\pi}\tan^{-1}(v_y/v_z)$ from the ES energy conserving algorithm (histogram), compared with analytical solutions of the Fokker-Planck equation (solid lines). Three snapshots in time are shown. Simulation results show excellent agreement with theoretical expectations. Sample size $N = 10^4$. }
    \label{fig:beam_dist}
\end{figure}

\begin{figure}
    \centering
    \includegraphics[width=0.8\linewidth]{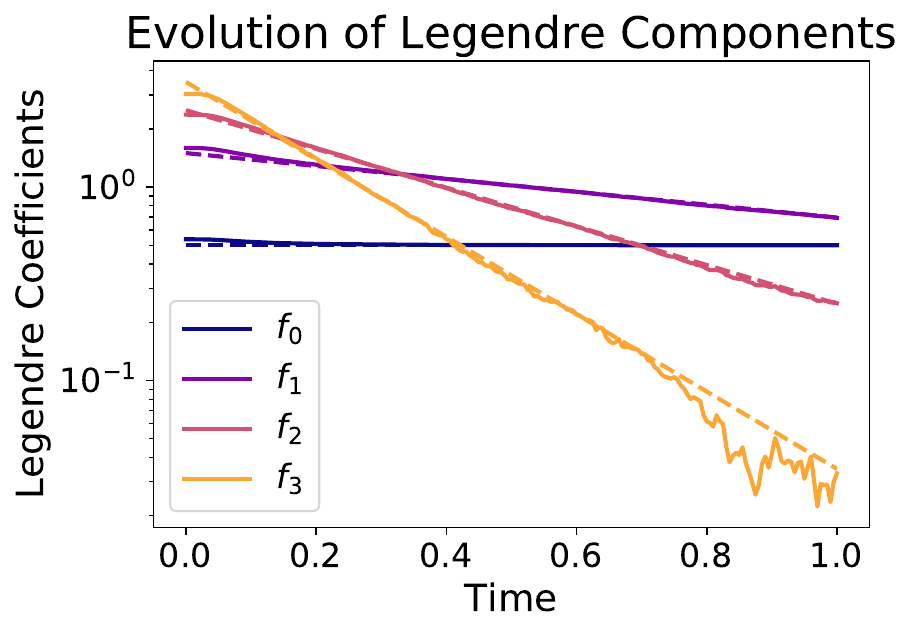}
    \caption{Simulated time evolution of coefficients $f_l$ for the Legendre components (solid lines) compared with analytical solution (dashed lines). The first 4 coefficients are shown ($>99\%$ of total distribution). Simulation results agree well with theoretical expectation. The increased noise for $f_4$ is attributed to finite-sampling noise. Sample size $N = 10^4$.}
    \label{fig:Fl}
\end{figure}

\begin{figure}
    \centering
    \begin{subfigure}[t]{0.8\linewidth}
        \centering
        \includegraphics[width=\linewidth]{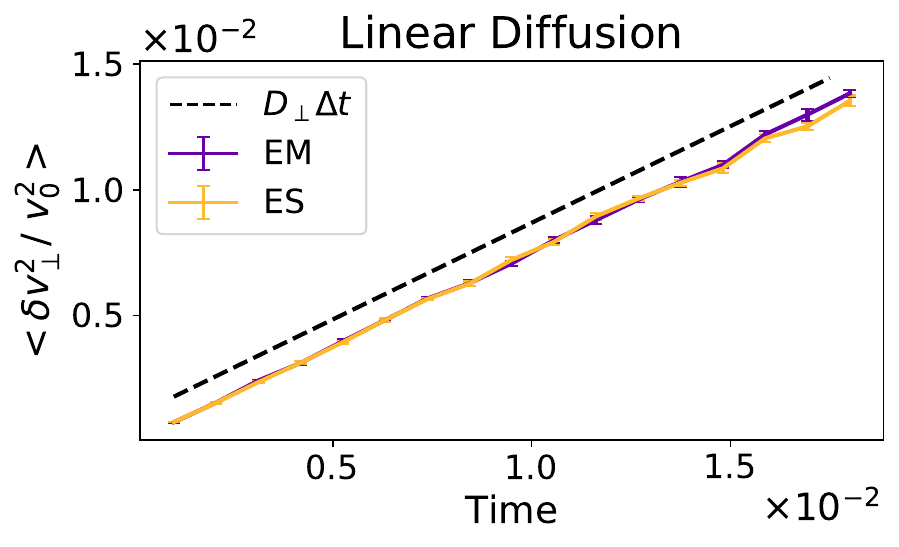}
    \end{subfigure}%
    
    \begin{subfigure}[t]{0.8\linewidth}
        \centering
        \includegraphics[width=\linewidth]{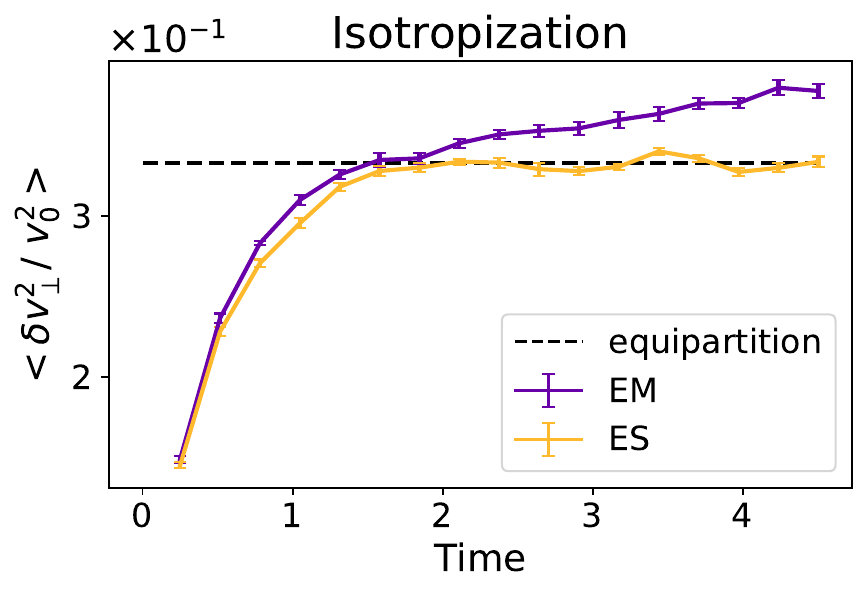}
    \end{subfigure}
    \caption{The short (top) and long (bottom) time behavior of the evolution of the normalized jump moment $\expval{\delta v_{\perp}^2/v_0^2}$. While the EM method results in numerical heating, the ES energy conserving algorithm show excellent long time accuracy and agreement with theoretical expectations (shown as black dashed lines). All time normalized to collision time. FP solution for short time limit is shifted upwards for visualization. Sample size $N = 10^3$}
    \label{fig:beam}
\end{figure}

The time evolution of $f$ can also be intuitively interpreted. Since higher orders of Legendre polynomial decay exponentially faster in time, the distribution will be ``smeared" out into a  uniform distribution in $\sin\theta$ as time goes on. In deed, when taking the long time limit $t \to \infty$, the only nonzero term left in the summation is $l=0$, indicating a uniform distribution on the $v = v_0$ sphere.

The particle distributions are calculated by integrating the Ito SDE Eq.~(\ref{eq:SDE_Ito}) with the traditional Euler-Maruyama (EM) method, and the Stratonovich SDE Eq.~(\ref{eq:SDE_Strat}) with the proposed explicitly solvable (ES) algorithm. Figure~\ref{fig:circle} shows the locations of simulated particles in velocity space $(v_\parallel, v_\perp)$. Without loss of generality, we chose the sign of $v_\perp$ to be the same as that of $v_y$. We can see that while the EM method produces a large spread in particle speed (the radial distance to the origin), the proposed ES method confined the particles exactly on the spherical surface. This is consistent with the previous numerical results shown in Fig.~\ref{fig:speed_rel}. In all figures, time is normalized with collision time $\tau_0 = 1/\nu_{ei}$.

The histograms of particle velocity distributions calculated by the ES algorithm are shown in Fig.~\ref{fig:beam_dist} as 3 snapshots in time. The corresponding analytical solutions from Eq.~(\ref{eq:f_solved}) are overlaid with the histograms. Only the first 10 terms are used in the Legendre series. Both the spectra of $v_\parallel$ and the azimuthal symmetry represented by the distribution in angle $\phi$ show excellent agreement with the analytical solution. The time evolution of the Legendre components is also shown in Fig.~\ref{fig:Fl}. The Legendre coefficients for the simulated distribution are found by fitting a truncated Legendre series to the histogram of velocity distribution. Slight deviations from theoretical expectation can be seen at small $t$ since higher order components have larger contributions at $t \to 0$ but are artificially truncated in the numerical fit.

The diffusion process can also be studied by calculating the second jump moment of the velocity distribution, shown in Fig.~\ref{fig:beam}. In the short time limit, both the EM and the ES method recovered the diffusion coefficient of the FP equation with minimal uncertainty. In the long time limit, we expect an isotropic distribution that becomes stationary in time as is evident from the FP solution~(\ref{eq:f_solved}). We can see from Fig.~\ref{fig:beam} that this limit is reached by the ES method after about 2 collision times, whereas the EM method continues to show clear numerical heating. This observation is also consistent with what is shown in figures \ref{fig:speed_rel} and \ref{fig:circle}, where the EM method injects a ``spread" in particle speed.

\section{Conclusion}
In this work, we present an energy-conserving numerical algorithm to integrate the Langevin equation for pitch angle scattering. Although the algorithm is formally implicit, it can be solved explicitly and is straightforward to implement. The algorithm converges globally with a similar order as that of the classic Euler-Maruyama method. However, since the velocity trajectories are confined to the sphere of constant speed, the numerical errors are effectively diverted to the azimuthal ``phase" in velocity space. This means that the dynamics of the distribution functions are not influenced by the build up of numerical errors, as can be seen from the beam diffusion example. Consequently, the proposed explicitly solvable algorithm is a good candidate for integrating the pitch angle scattering operator because of its excellent long time accuracy.

Future work may include implementing the ES method under various external electromagnetic fields and studying its numerical properties. We also aim to generalize the ES method to more complex collision operators, and to apply the algorithm in more realistic plasma physics problems. An extension to solving nonlinear Fokker-Planck equations using SDEs that depend explicitly on distribution functions is also possible \cite{barbu2020nonlinear, wang2018distribution, allen1994computational, frank2005nonlinear}. 

\begin{acknowledgments}
X. Zhang would like to thank Francesca Poli and Nicolas Lopez for helpful suggestions. H. Qin thanks Tom Tyranowski and Yajuan Sun for fruitful discussions. This work is supported by DOE contract number {DE-AC02-09CH11466}. The digital data for this paper can be found
at \url{http://arks.princeton.edu/ark:/88435/dsp011v53k0334}.
\end{acknowledgments}

\appendix
\section{Catastrophic Drift in Energy}\label{sec:app}
 Stochastic calculus in general is strongly coupled to the choice of numerical schemes. For example, in deterministic calculus, the forward and backward Euler integration of the differential equation
\begin{equation}
    \pdv{f}{t} := f'(t) = C f(t)
\end{equation}
will inevitably converge to the same result as the step size $h$ approaches zero. However, such is not the case in stochastic calculus. 

If one were to erroneously integrate a Stratonovich SDE with the Euler-Maruyama method, for example, the equation being integrated numerically ends up being a different SDE. Since a Stratonovich can be converted to an Ito SDE, and vice versa, via the relation
\begin{align}
    a(t, x)dt + b(t, x)\circ dW_t =
                \tilde{a}(t, x)dt + b(t, x) dW_t,
\end{align}
where 
\begin{equation}
    \tilde{a}(t, x) = a(t, x)dt+ \frac{1}{2}b(t, x)\pdv{b}{x}(t, x),
\end{equation}
the error in choosing the correct numerical scheme will lead to a spurious drift, which could be at the same order of magnitude as the actual drift or the variable $X_t$ itself.

\begin{figure}[h]
    \centering
    \includegraphics[width=0.8\linewidth]{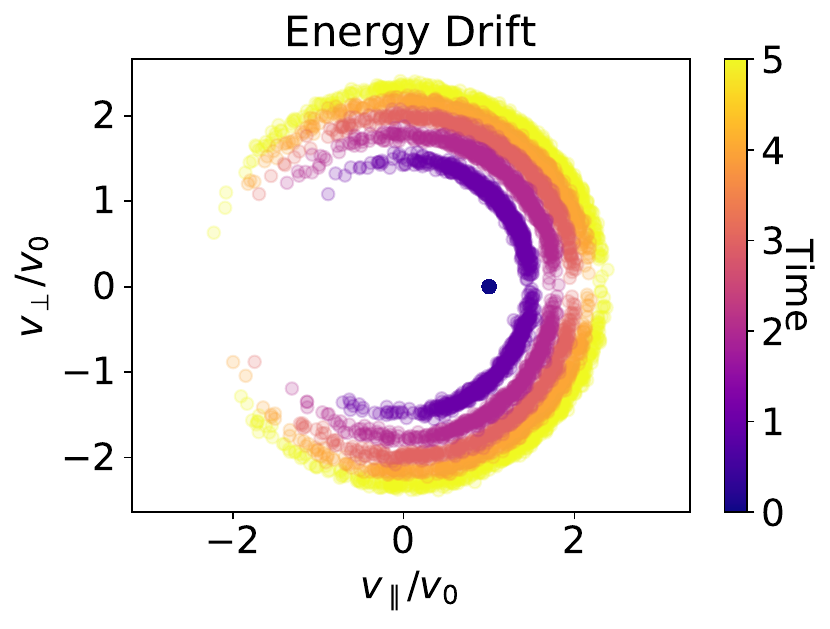}
    \caption{Catastrophic drift in energy resulting from integrating the Stratonovich SDE with the Euler-Maruyama method. After 5 collision times, the average speed of particles have doubled. Five snapshots in time are shown. Sample size $N = 10^3$.}
    \label{fig:app}
\end{figure}

As a simple example, we integrate equation~(\ref{eq:SDE_Strat}) with the Euler-Maruyama method Alg.~\ref{alg:EM}. The resulting particle distribution in velocity space is shown as 5 snapshots in time, in figure~\ref{fig:app}. We can see that the radius of the circle that the velocity vectors lie on, which corresponds to the speed of the particles, is steadily increasing in time. By 5 collision times, the average speed of the particles have almost doubled. This is quite an unacceptable result.

\section{Global Strong Convergence}\label{sec:proof}
For a given current state $\vnow$, assume the exact solution for next step is $\vv_{k+1}$. The one-step difference between the ES method Eq.~(\ref{eq:numerical_SDE_ES}) and EM method Eq.~(\ref{eq:numerical_SDE_EM}) at next step is $\bar{\vv}^{ES}_{k+1} - \bar{\vv}^{EM}_{k+1}$:
\begin{align}
& \dfrac{1}{2}\vb{M}_k\times (\bar{\vv}^{ES}_{k+1} - \vv_{k}) + \dfrac{\vv_k}{v_k^3} h \nonumber \\
=& \dfrac{1}{2}\vb{M}_k\times \left(\vb{M}_k\times \bar{\vv}_{k+1/2}\right) + \dfrac{\vv_k}{v_k^3} h,
\label{eq:difference}
\end{align}
where we have defined $\vb{M}_k:=\vv_k\times \Delta \W/v_k^{5/2}$ for convenience of notations.

Notice that for the theorem on the strong convergence in \cite{milstein2013stochastic}, the order of strong convergence is defined in the mean-square sense. Thus in this section, we used the following two definition of strong error:
\begin{align}
\epsilon_\text{s}^2 &:= 
\left\langle\left|\vv(T;t_0,\vv_0)-\bar{\vv}(T;t_0,\vv_0, h)\right|^2\right\rangle;\\
\epsilon_\text{s,o}^2 &:=  \left\langle\left|\vv(t_0+h; t_0,\vv_0)-\bar{\vv}(t_0+h; t_0,\vv_0,h)\right|^2\right\rangle.
\label{eq:strong_ms}
\end{align}

Due to the Lyapunov inequality \cite{kloeden2013numerical}:
\begin{align}
\langle|X|\rangle \leq \sqrt{\langle |X|^2\rangle},
\end{align}
the strong error we defined in absolute sense in section \ref{sec:Strong_and_weak} is bounded by the root-mean-squared error, and therefore has convergence rate up to that demonstrated here.

Firstly we estimate the one-step strong error. Using the triangle inequality: 
\begin{align}
\begin{split}
\langle|X+Y|^2\rangle & \leq 2 \langle|X|^2\rangle + 2 \langle|Y|^2\rangle \\
& \sim \mathcal{O}(\langle|X|^2\rangle) + \mathcal{O}(\langle|Y|^2\rangle),
\end{split}
\end{align} 
we can estimate the one-step strong error of ES method by:
\begin{align}
& \left\langle|\bar{\vv}^{ES}_{k+1} - \vv_{k+1}|^2 \right\rangle \nonumber  \\
= &  \left\langle|\bar{\vv}^{ES}_{k+1} - \bar{\vv}^{EM}_{k+1} + \bar{\vv}^{EM}_{k+1} - \vv_{k+1}|^2 \right\rangle \nonumber \\ 
\begin{split}
\sim &\, \mathcal{O}\left(\left\langle|\bar{\vv}^{EM}_{k+1} - \vv_{k+1}|^2\right\rangle\right)+ \mathcal{O}\left(\left|\dfrac{\vv_k}{v_k^3} h\right|^2\right)\\
& + \mathcal{O}\left(\left\langle\left|\dfrac{1}{2}\vb{M}_k\times \left(\vb{M}_k\times \bar{\vv}_{k+1/2}\right)\right|^2\right\rangle\right).
\label{eq:estimate_s}
\end{split}
\end{align}

The first term in Eq.~(\ref{eq:estimate_s}) is the one-step strong error of the EM method and is known to be $\mathcal{O}(h^2)$ \cite{milstein2013stochastic}. The second term is deterministic and is of order $\mathcal{O}(h^2)$. Since the expected norm of the Wiener function is $\langle|{\delW}^2|\rangle\sim\mathcal{O}(h)$, the expectation for the size of $\vb a_k$ is also of the same order $\langle|{\vb{M}_k}^2|\rangle\sim\mathcal{O}(h)$. The third term in Eq.~(\ref{eq:estimate_s}) therefore also scales as $\mathcal{O}(h^2)$. Thus the one-step strong error of ES method is of order 1. 

Next we estimate the one-step weak error of the ES method. Using Eq.~(\ref{eq:numerical_SDE_ES}), we have:
\begin{align}
\bar{\vv}_{k+1/2} = \bar{\vv}^{ES}_{k+1} + \dfrac{1}{2} \mathbf{M}_k \times \bar{\vv}_{k+1/2}.
\label{eq:v_1_2}
\end{align}
Plugging the equation above into Eq.~(\ref{eq:difference}), we obtain the difference between ES and EM as:
\begin{align}
&  \dfrac{1}{2}\vb{M}_k\times \left(\vb{M}_k\times \bar{\vv}_{k+1/2}\right) + \dfrac{\vv_k}{v_k^3} h  \nonumber\\
\begin{split}
= &   \dfrac{1}{2}\vb{M}_k\times \left(\vb{M}_k\times {\vv}_k \right) + \dfrac{\vv_k}{v_k^3} h 
\\
& +  \dfrac{1}{8}\vb{M}_k \times [\vb{M}_k 
\times \left( \vb{M}_k\times \bar{\vv}_{k+1/2} \right)].
\end{split}
\end{align}

Applying the triangle inequality yields:
\begin{align}
& \left|\langle\bar{\vv}_{k+1}^{ES}\rangle - \langle\vv_{k+1}\rangle\right| \nonumber \\
\leq & \left|\langle\bar{\vv}_{k+1}^{ES}\rangle - \langle\bar{\vv}^{EM}_{k+1}\rangle\right|
+ \left|\langle\bar{\vv}_{k+1}^{EM}\rangle - \langle{\vv}_{k+1}\rangle\right| \nonumber \\
\begin{split}
\leq & \left|\langle\bar{\vv}_{k+1}^{EM}\rangle - \langle{\vv}_{k+1}\rangle\right| \\
& +  \left| \left\langle\dfrac{1}{2}\vb{M}_k\times \left(\vb{M}_k\times {\vv}_k \right)\right\rangle + \dfrac{\vv_k}{v_k^3} h \right| \\
& + \left| \left\langle \dfrac{1}{8}\vb{M}_k \times [\vb{M}_k 
\times \left( \vb{M}_k\times \bar{\vv}_{k+1/2} \right)] \right\rangle\right|.
\label{eq:estimate_w}
\end{split}
\end{align}

The first term in Eq.~(\ref{eq:estimate_w}) is the one-step weak error of EM method, which is known to be $\mathcal{O}(h^2)$ \cite{milstein2013stochastic}. Because of the double cross product and the fact that $\vb{M}_k\cdot \vv_k=0$, we find that the expectation in the second term in Eq.~(\ref{eq:estimate_w}) cancels out the deterministic term exactly:
\begin{align}
\left\langle \dfrac{1}{2}\vb{M}_k\times \left(\vb{M}_k\times {\vv}_k \right) \right\rangle = - \dfrac{\vv_k}{v_k^3} h.
\end{align}
Using $\langle{|\vb{M}_k|}\rangle\sim\mathcal{O}(h^{1/2})$ again, we see that the third term in Eq. (\ref{eq:estimate_w}) is at most $\mathcal{O}(h^{3/2})$. So the one-step weak error of the ES method $|\langle\bar{\vv}_{k+1}^{ES}\rangle - \langle\vv_{k+1}\rangle|$ is also at most of order $\mathcal{O}(h^{3/2})$. Therefore, the ES method also has order $1/2$ global strong convergence, same as the Euler-Maruyama method.

\section{Explicit Form for ES Velocity Update}\label{sec:explicit_cayley}

The explicit form of the Cayley transform is derived as follows. The vector $\vb{M}$ in Eq.~(\ref{eq:cross}) can be explicitly given as:
\begin{align}
\vb{M} = \begin{pmatrix}M_x \\ M_y \\ M_z\end{pmatrix} = \dfrac{1}{2 v^{5/2}}
\begin{pmatrix}
v_y \Delta W_z - v_z \Delta W_y \\
v_z \Delta W_x - v_x \Delta W_z \\
v_x \Delta W_y - v_y \Delta W_x \\
\end{pmatrix},
\end{align}
where subscript $k$ (for time steps) is omitted for simplicity of notations. It is easy to calculate that the norm-squared of the vector $\vb{M}$ is
\begin{align}
M^2=\vb{M}\cdot\vb{M}= \left[ v^2 \Delta W^2 - (\mathbf{v}\cdot \Delta \boldsymbol{W})^2 \right] / 4 v^5.
\end{align}

Due to its skew-symmetric nature, the following two identities hold for $\hat{\vb{M}}$:
\begin{align}
\hat{\vb{M}}^2&=-M^2 \vb{I}+\vb{M}\vb{M},\label{eq:skew1}\\
\hat{\vb{M}}^3&=-M^2\hat{\vb{M}}, \label{eq:skew2}
\end{align}
where $\vb{M}\vb{M}$ is the tensor product of vector $\vb{M}$. Using Eq.~(\ref{eq:skew2}), we find that:
\begin{align}
(\vb{I}-\hat{\vb{M}})^{-1} = \vb{I} + \dfrac{1}{1+M^2}(\vb{M}+\vb{M}^2).
\end{align}
Thus the Cayley transform defined in Eq.~(\ref{eq:cayley}) can be simplified as:
\begin{align}
\begin{split}
\mathcal{C}(\hat{\vb{M}}) & :=(\vb{I}-\hat{\vb{M}})^{-1}(\vb{I}+\hat{\vb{M}}) \\
& = \vb{I} + \dfrac{2}{1+M^2}(\vb{M}+\vb{M}^2) \\
& = \dfrac{1}{1+M^2} \left[ (1-M^2)\vb{I} + 2\hat{\vb{M}} + 2\vb{M}\vb{M} \right]
\end{split}
\end{align}

Noticing that,
\begin{align}
\vb{MM}\cdot \vb{v} 
= \vb{M}(\vb{M}\cdot\vb{v})
= \vb{M} \dfrac{(\vb{v}\times \Delta \boldsymbol{W})\cdot \vb{v}}{2v^{5/2}} = 0,
\end{align}
The one-step approximation $\bar{\vb{v}}$ is therefore given explicitly by:
\begin{align}
\begin{split}
\bar{\vb{v}} &= \mathcal{C}(\hat{\vb{M}}) \vb{v} \\
& = \dfrac{1}{1+M^2} \left[ (1-M^2)\vb{I} + 2\hat{\vb{M}} \right] \vb{v} \\
& = \dfrac{1}{1+M^2} \left[ (1-M^2)\vb{v} + 2\vb{M}\times \vb{v} \right].
\end{split}
\label{eq:ES_explicit}
\end{align}

In addition, with this explicit form, we can easily verify that the one-step approximation conserves the magnitude of the velocity:
\begin{align*}
\bar{\vb{v}} \cdot \bar{\vb{v}} = \dfrac{(1-M^2)^2 v^2 + 4 (\vb{M}\times \vb{v})\cdot(\vb{M}\times \vb{v})}{(1+M^2)^2} = v^2.
\end{align*}

From Eq.~(\ref{eq:numerical_SDE_ES}) it is clear that the direct discretization of the Stratonovich form of the SDE (\ref{eq:SDE_Strat}) using the mid-point method conserves energy, and the direct discretization of the Ito form (\ref{eq:SDE_Ito}) using the Euler-Maruyama method does not. However, this does not imply that the Ito form does not admit energy-preserving discretization. In fact, since our algorithm is explicitly solvable, it is possible to transform  Eq.~(\ref{eq:numerical_SDE_ES}) into a discretization of the Ito SDE Eq.~(\ref{eq:SDE_Ito}). Eq.~(\ref{eq:ES_explicit}) could be written as:
\begin{align}
\bar{\vb{v}} - \vb{v} = 
-\dfrac{2M^2}{1+M^2}\vb{v}
+\dfrac{1}{1+M^2}\dfrac{(\vb{v}\times \Delta \boldsymbol{W})\times \vb{v}}{v^{5/2}},
\end{align}
which is similar to the Euler-Maruyama method in Eq.~(\ref{eq:numerical_SDE_EM}) but has modified drift and diffusion coefficients. This can be viewed as an energy-preserving algorithm for the Ito SDE.

\bibliography{Coulomb}

\end{document}